# Parallel Optimisation of Bootstrapping in R

T. M. Sloan, M. Piotrowski
EPCC, University of Edinburgh
Edinburgh, UK
t.sloan@epcc.ed.ac.uk

T. Forster, P. Ghazal
Division of Pathway Medicine, University of Edinburgh
Edinburgh, UK

*Abstract*—Bootstrapping is a popular and computationally demanding resampling method used for measuring the accuracy of sample estimates and assisting with statistical inference. R is a freely available language and environment for statistical computing popular with biostatisticians for genomic data analyses. A survey of such R users highlighted its implementation of bootstrapping as a prime candidate for parallelization to overcome computational bottlenecks. The Simple Parallel R Interface (SPRINT) is a package that allows R users to exploit high performance computing in multi-core desktops and supercomputers without expert knowledge of such systems. This paper describes the parallelization of bootstrapping for inclusion in the SPRINT R package. Depending on the complexity of the bootstrap statistic and the number of resamples, this implementation has close to optimal speed up on up to 16 nodes of a supercomputer and close to 100 on 512 nodes. This performance in a multi-node setting compares favourably with an existing parallelization option in the native R implementation of bootstrapping.

## I. Introduction

The statistical programming language R [1] is a highly popular, free software environment commonly used for data analysis, including biostatistics and bioinformatics, of very large data sets. The Simple Parallel R INTerface (SPRINT) R package contains parallel implementations of key functions of use to many such analyses [4] in order to provide R users with an easy route to exploiting High Performance Computing on multi-core desktops, supercomputers and clouds [15].

SPRINT provides drop-in replacements for a number of computationally expensive R functions that were identified as important in a user requirements survey of the bioinformatics community [2,12]. These drop-in replacements are parallelized using the Message Passing Interface (MPI) [20], with data distribution carried out transparently from the end-user's point of view [12]. For a more detailed description of the SPRINT architecture, see [14]. Users of SPRINT write R analysis scripts as they would have performed previously for serial analysis [12]. To compute in parallel, the R script must be executed like any other parallel MPI task (e.g. mpiexec -n 16 R -f script.R) [12].

The survey [2] of the R user community by the SPRINT team identified those R functions causing computational bottlenecks in the processing of genomic data or seen as intractable on desktop machines. Respondents were asked to list the five R functions they consider most useful for inclusion in a parallel R function library. Bootstrapping [3] was the 3rd most requested function. Bootstrapping is a very generic function with applicability wherever estimates or results are calculated on data.

This paper focuses on the development of a parallel implementation of bootstrapping for inclusion in the SPRINT R package. The paper includes the results from an investigation into the performance of this implementation.

## II. Bootstrapping

Bootstrapping, introduced by Efron in the late 1970's, is strongly linked with the development of computing systems capabilities [6]. It allows robust estimation of any sort of estimate and result that can be calculated on data. Some examples are mean, median, standard deviations, ratios, differences, hypothesis tests, complex equations and clustering. Effectively anything that can be calculated once, can then be bootstrapped and calculated thousands of times to get a "better" result. It is often used in situations where:

- the distribution of a statistic is unknown,
- the sample size is not large enough for statistical inference, or,
- when only a small sample of a larger population is available [5].

If the distribution of a statistic is unknown, bootstrapping provides a mean of assessing the properties of the distribution [6]. If a sample size is not large enough to determine statistical inference but the distribution is known then bootstrapping can be used to account for distortions of a small sample that may not be representative of the population [6]. In studies where only a small sample is available of a larger population, bootstrapping can be used to estimate the variance of the population [6]. Bootstrapping is performed by measuring a property on a number of generated samples where each sample is created such that it may have been present in the original dataset [6].

Bootstrapping has proven to be a popular technique and has been applied to a wide range of applications outside of genomic data analysis.

Bootstrapping is generally used to determine uncertainty in population estimation, replacing mathematical analysis with computer simulation. On the original observation data set, resampling with replacement is applied to generate new sample data sets with an equal number of values as the original data set, but highlighting different distribution properties of the original data. A higher number of resample data sets improves the accuracy of the final estimation, but requires more computational processing power. In the current era of data intensive research, the number of bootstrap resamples i.e. the resample data sets, is usually limited by the available computing resources.

An example usage of bootstrap is where one wishes to estimate the mean $M$ from an unknown population $P$ on the basis of randomly sampled data. Bootstrap is applied to this as follows.

1. Calculate the sample mean $m$. Now need to estimate the standard error of $m$ to assess the amount of uncertainty in our estimate. This depends on the variance of our unknown population $P$. So,

2. simulate the entire population distribution using just the sample provided by assuming that both the sample and population have a similar shape.

3. Now generate new samples by resampling the original with replacement. This introduces variability into our measurements.

4. Estimate the variance of $P$ by calculating the standard deviation of the multiple $m$ values obtained from the new samples.

### III. R AND BOOTSTRAPPING

In R, the `boot` function in the package boot [16], allows users to generate bootstrap samples. It can bootstrap any statistical function that can be expressed in R and from these samples it can generate estimates of bias and bootstrap confidence intervals. The boot function executes the resampling of a specified dataset and calculates the specified statistic of interest on these samples.

Below is an example usage of bootstrapping in R. This is based on the documentation and data supplied with the R boot package [16].

The statistic to be bootstrapped is defined as the ratio function and it is the 49 U.S. cities population increase ratio between 1920 and 1930. This is how we define the statistic:

```
> library(boot)
> ratio <- function(d, w) sum(d$x * w)/sum(d$u * w)
```

The first line contains the instruction for loading the boot package into R. The second line defines a function to be called `ratio`. In this definition, the first argument passed to `ratio` is the dataset `d`. The function expects this dataset to contain the columns `x` and `u` where in R terminology they are referred to as `d$x` and `d$u` respectively. The second argument, `w`, is an index vector of the observations in the dataset to use or a frequency or weight vector that informs the sampling probabilities. This example uses the vector of importance weights.

Now we can call the boot command as follows.

```
> boot(bigcity, ratio, R = 999, stype = "w")
```

In this call the data set `bigcity` is passed as well our statistic function, `ratio`, the number of resamples R (also known as replicates) to produce, and the statistic type, `stype="w"`, to indicate that the second argument of ratio is the vector of importance weights. This vector is generated by `boot`.

Within `boot`, the actual execution of the bootstrapping is broken down into three steps [6]:

1. creating random indices to produce the resample data sets

2. executing the statistic on the resample data sets

3. calculating results.

Crucially, the number of bootstrap resamples is related to the random sampling error. This error decreases with increasing bootstrap resamples. Computational efficiency is therefore key in reducing the error in any estimated parameters. As a consequence, researchers often choose to carry out only a few hundred resampling steps in order to reduce their computation time to acceptable levels ([7]-[9]) for their computational infrastructure. This in effect limits their science by trading accuracy for performance. This especially applies to high-dimensional (e.g. microarray based technology) data with thousands of variables, where this problem is multiplied.

### IV. PREVIOUS WORK

In a previous prototype SPRINT parallelization of bootstrapping for R [6] in 2010, only the 2$^{nd}$ step, executing resamples, was parallelized in a simplistic manner and this prototype could not be invoked from R in the way a user expects. This prototype parallelization achieved a speed up of only 8 on 16 cores with a synthetic dataset [6] on a small shared memory computational platform due to the restrictions of the original SPRINT architecture [4]. In this architecture only one process could interact with the R runtime environment. This prototype was therefore not able to benefit from accessing the R interpreter simultaneously on all processes involved in computation. As the result it was not compatible with the rest of the parallel functions in the SPRINT package and hence could not be combined in parallel workflows. It also implemented only a fraction of the original bootstrap functionality being limited only to non-parametric standard simulation.

A synthetic dataset was used in benchmarking on this small platform due to the long elapsed times when running typical genomics datasets consisting of approximately 22,000 genes from hundreds of patients [6]. This problem will be exacerbated with technology platforms in biology that now allow measurement of $10^5$ to $10^6$ exons, Single Nucleotide Polymorphisms (SNPs) or "short reads" from the genome. Bootstrap has been applied (usually again with hundreds rather than thousands of bootstrap resamples) to identify SNPs in

RNA-Seq data [10] or in the assessment of read distributions for exons in RNA-Seq data [11].

## V. PBOOT - SPRINT PARALLEL IMPLEMENTATION OF BOOTSTRAPPING IN R

Following previous SPRINT-related work on random forest classification and rank product statistical tests [12] in R, more efficient techniques for random number generation and result combinations are now available. These provide better performance and enable the SPRINT user interface of the parallelized bootstrapping to be as the R user would expect. The specific areas optimized are in the use of more efficient serialization and de-serialization, the replacement of indexed sends since these are linear in nature and the use of built-in R functions for processing of the statistic. Moreover the data distribution and gathering of the statistic results now use a tree-like rather than linear approach. This tree-like approach was particularly successful in the SPRINT implementation of random forest in R [12]. It should be noted that once the data and statistic to be used have been distributed, there is little if any communication required in bootstrapping until the final results gather.

The latest SPRINT architecture allows the spawning individual R processes and their associated runtime environment on all computational nodes [14]. Parallelization of appropriate R functions can then achieved by data distribution, that is, each parallel process executes the same R function on a fraction of the data set. This approach can only be applied in situations where no data dependency occurs such as in bootstrapping. This significantly simplifies the parallelization of R functions that are written partially or purely in R. These no longer have to be rewritten in C, but instead can be serialised and sent to all the processes where they can be deserialised and passed back to the R interpreter that carries out the computation.

This model also allows the sending of more complex R objects between the processes, as well as user defined functions that are passed as arguments. This approach ideally suits the SPRINT parallel bootstrap implementation, `pboot()`, where a copy of a given statistic function is executed on each process using a number of different samples of the data. Here there is no data dependency between each process and the bootstrapped statistic is not known in advance and thus cannot be rewritten in C.

Furthermore, this approach means `pboot` can have an almost identical interface to the serial version, `boot`. In `pboot`, all the function's arguments are serialized and used when executing the statistic on the number of resamples on all the participating nodes.

The process of creating new indices, i.e. step 1 of bootstrapping, is currently performed in the serial part of `pboot` and then the new replicates are redistributed to all the workers. This definitely has an impact on the performance and scalability of the parallel implementation, but was chosen for simplicity and to allow, when appropriate, reproducibility of results when compared with the sequential version.

Step 2 of bootstrapping, executing the statistic on the resamples is parallelized. In this, at the end of an iteration, a single bootstrap statistic is calculated on each process and added to a local list.

After all computation is finished, these lists are combined using the SPRINT parallel reduction function as part of step 3 of the bootstrapping, i.e. calculating results. This parallel reduction was initially implemented to speed up the tree reduction algorithm in SPRINT's parallel random forest implementation. Here, a combine function can be passed as an argument to the parallel reduction function as long as it is associative. This parallel reduction algorithm can concatenate the result lists with logarithmic rather than linear complexity, thus improving the performance significantly.

The `pboot` user interface is almost identical to its sequential counterpart, `boot`. In an existing user's R script, the minimal changes required to run it include loading the SPRINT library, renaming `boot` function call to `pboot` and terminating SPRINT at the end of the script via the `pterminate` function call. By way of illustration, here is the previous bootstrapping example now performed using `pboot`.

```
> library(boot)
> library(sprint)
> ratio <- function(d, w) sum(d$x * w)/sum(d$u * w)
> pboot(bigcity, ratio, R = 999, stype = "w")
> pterminate()
> quit()
```

## VI. METHODS

The performance results were gathered on the Phase 3 system of HECTOR, the UK's national supercomputing service [17]. This is a CRAY XE supercomputer comprising 2816 computing nodes, each containing two 16-core AMD Opteron 2.3Ghz Interlagos processors, which gives a total number of 90,112 available cores. This runs the Linux operating system. R Version 2.15.2, SPRINT Version 1.0.4 and boot Version 1.3-7 were used in the benchmarks.

All benchmark runs were performed using data from Golub [18] that is available as an R package. These data are the combined training samples and test samples. There are 47 patients with acute lymphoblastic leukemia (ALL) and 25 patients with acute myeloid leukemia (AML), data on the expression of 7129 genes are available.

Benchmarks were collected to:

- investigate the impact of the complexity of the bootstrap statistic on performance,
- investigate the dependency of performance on the number of resamples,
- compare performance with a native R parallel boot implementation.

We have performed bootstrap using two different statistical functions, median and standard deviation, to investigate how scalability and efficiency depends on the bootstrap statistic and the number of resamples. The scalability and efficiency have been calculated relative to the serial performance of the existing implementation of `boot()` in boot version 1.3-7.

Since 2011, the `boot` function, in the R package boot, has been extended to exploit parallel computing thus enabling researchers to increase the number of resampling steps executed in acceptable computation times. Version 1.3-7 of the boot package makes use of the R parallel package that has been available since R Version 2.14 as part of the R core. On a POSIX-based operating system such as Linux or Mac OS X, after loading the parallel package via the instruction

```
> library(parallel)
```

the call to the `boot` function in our example becomes

```
> boot(bigcity, ratio, R = 999, stype = "w", parallel
= "multicore", ncpus = 4 )
```

where `parallel="multicore"` specifies that the parallelization mechanism to be used is based on the POSIX `fork()` system call. This mechanism is applicable only to multiprocessor or multicore computers and cannot be used on clusters. In this call `ncpus = 4` indicates the number of computational processes to use to reduce the execution time of the `boot` function. This is typically set to the number of cores or processors available.

On Windows-based systems and Linux clusters, the call to the `boot` function instead becomes

```
> boot(bigcity, ratio, R = 999, stype = "w", parallel
= "snow", ncpus = 4 )
```

where `parallel="snow"` specifies that parallelization is based on socket connections. With this specification, alternative parallelization mechanisms can be used, eg. MPI [20] or PVM [21], but these require prior installation and configuration and installation of the relevant R packages eg. Rmpi and rpvm.

Of these parallel options in Version 1.3-7 of the boot package, only the `parallel= "multicore"` option was available on the HECTOR supercomputing service. In this service, this option was restricted to using a single node of the CRAY XE due to the option's reliance on the POSIX `fork()` system call. Similar, to the SPRINT `pboot()`, this option parallelizes step 2 of the bootstrap method, the execution of the statistic on the resamples. To compare the performance of `pboot()` with this native R parallelization, we also ran benchmarks of this with the same Golub dataset with the same bootstap statistic and number of resamples.

Finally, with the increasing volumes of data expected from next generation sequencing, the number of resamples used when biostatisticians apply bootstrap will increase. With that in mind, both boot with the multicore option and SPRINT pboot were tested to determine the maximum number of resamples they could cope with compared to serial boot. These tests were again undertaken on HECTOR with the same Golub dataset and with median as the bootstrap statistic.

## VII. RESULTS AND DISCUSSION

### A. Complex bootstrap statistics benefit from multi-node systems

Fig.1 shows the parallel speed up and efficiency when the bootstrap statistic is the median and the number of resamples is 24,999. The figure shows that when `pboot()` utilises 1 process per node, the performance for 2 to 16 nodes is close to linear with an efficiency of 90% or greater. At 16 nodes, the speed up is greater than 14. Between 32 and 128 nodes the rate of increase for both speed up and efficiency dips, however at 64 nodes, the speed up is still over 40 and at 128 it is over 55 with an efficiency above 40%. This dip is related to the overhead of data distribution and results gathering versus the amount of work in processing this size of dataset for this number of resamples.

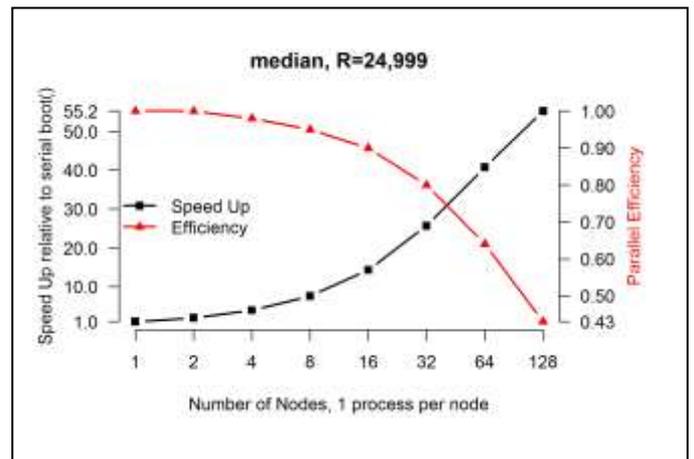

Fig. 1. Speed up and effficiency of pboot with median as the bootstap statistic and 24,999 resamples.

Fig.2 shows the parallel speed up and efficiency when standard deviation is the bootstrap statistic. The same data set and the same number of resamples are used as in the median benchmarks shown in Fig.1. Fig.2 shows that for standard deviation the scalability and efficiency are markedly less compared to when median is the bootstrap statistic. In Fig.2 between 2 and 16 nodes, the parallel efficiency goes from 97% to 80%, and the speed up at 16 nodes is just under 13. At 32 nodes and above, the differences in performance are more obvious, with a fall in the increase of speed up and efficiency such that by 128 nodes, speed up is around 36 and efficiency is around 30%. The results from Fig.1 and Fig.2 therefore show, not surprisingly, that the complexity of the bootstrap statistic will impact on the parallel performance that the SPRINT `pboot()` can achieve. That is, the greater the computational complexity of the statistic implementation, the better the likely performance gains when the work is spread across increasing number of nodes.

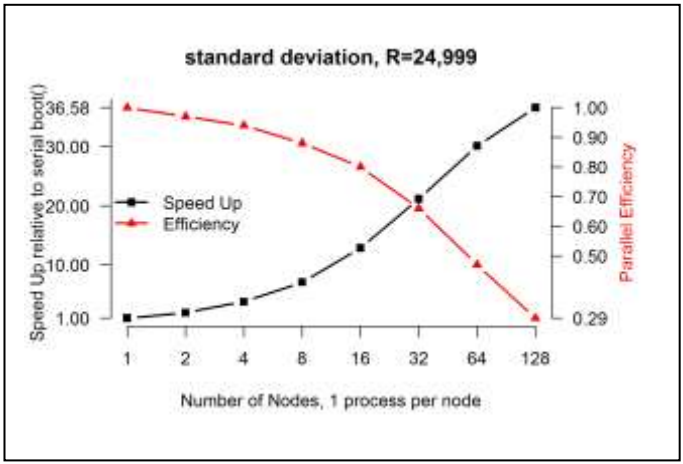

Fig. 2. Speed up and effficiency of pboot with standard deviation as the bootstap statistic and 24,999 resamples.

*B. Dependency of performance on number of resamples*

Fig.3 shows the parallel efficiency on the benchmark runs with standard deviation again as the bootstrap statistic but this time with a smaller number of resamples, 9,999 in this case. Here at 16 nodes, the speed up is under 12 and the efficiency has fallen to around 70%. At 32 nodes and above, the scalability and efficiency achieved are less than in the case of 24,999 resamples. At 128 nodes, the speed up is 27 and efficiency around 20%. Similar to the dependency on the bootstrap statistic, the results in Fig.3 shows that the performance gains from increasing nodes are dependent on the number of resamples required.

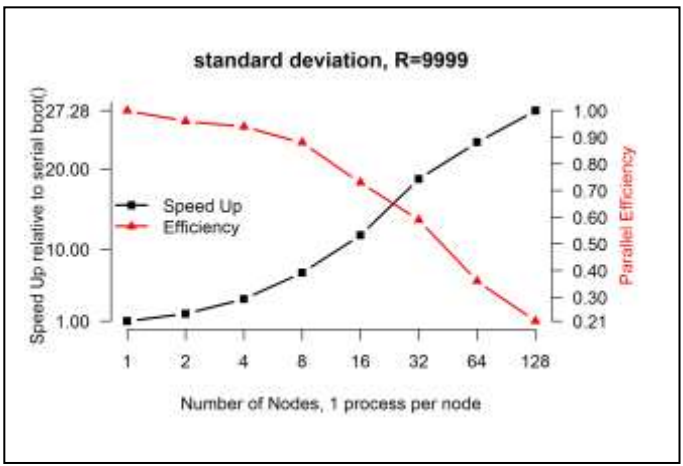

Fig. 3. Speed up and efficency of pboot with standard deviation as the bootstrap statistic and 9,999 resamples.

*C. SPRINT pboot circumvents native R parallelization cluster limitations*

Fig.4 shows the speed up of the multicore option along with SPRINT `pboot()` when all the processes run on a single node and when each process runs on its own node i.e. 1 process per node. These benchmarks were run on HECTOR with the same Golub dataset, median as the bootstrap statistic and 24,999 as the number of resamples. The figure shows that at low number processes, ie 2, 4 and 8, the speed up of pboot on multiple nodes (i.e. 1 process per node) and the multicore option are comparable At 16 processes and more obviously at 32 processes, the speed-up for pboot is greater. Above 32 processes, the multicore option cannot run on HECTOR due to its restriction of 1 process per core.

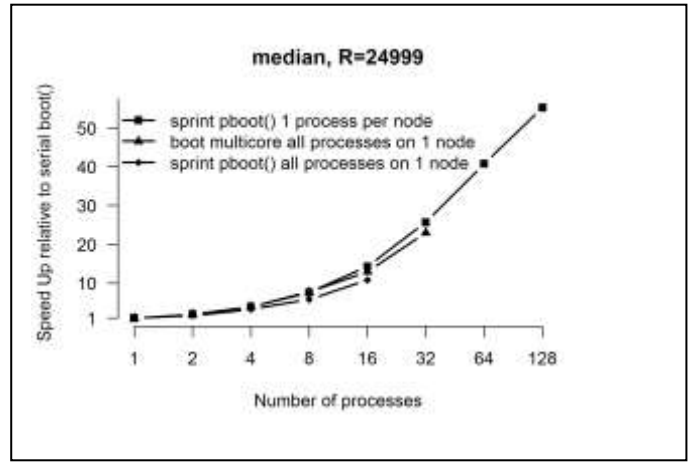

Fig. 4. Speed up of boot with multicore option compared to pboot with median as the bootstrap statistic and 24,999 resamples.

*D. SPRINT pboot provides advantages on multi-node systems, native R parallelization provides advantanges on single-node systems*

Fig. 4 shows the speed up for pboot when all the processes are on a single node. The speed up for this is markedly less and can only run on a maximum of 16 processes on HECTOR. This is due to the memory overhead of the SPRINT architecture where as previously mentioned in this paper, a separate instance of R is invoked by each participating process. This overhead is more apparent in Fig.5 which shows the parallel efficiency of these benchmarks. This clearly shows that when all the processes are run on a single node then the multicore option makes more efficient use of the parallel resources compared to pboot when all the processes are on 1 node.

At 16 and 32 processes, however, this figure also shows how pboot, with the processes on multiple nodes (i.e. 1 process per node), is in turn more efficient than this multicore option.

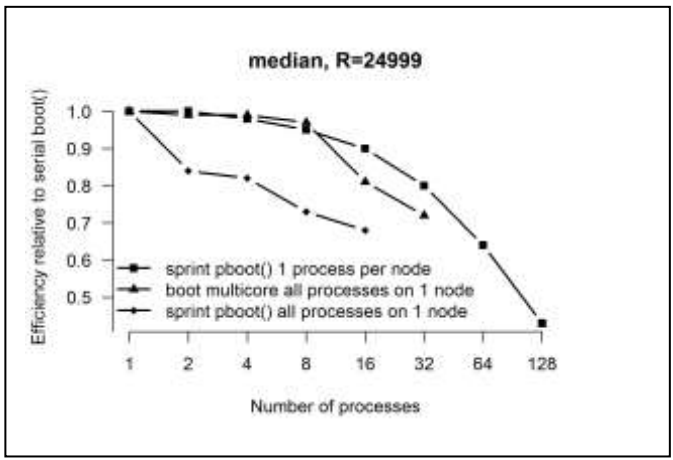

Fig. 5. Parallel efficiency of boot with multicore option compared with pboot.

*E. Serial boot allows more resamples than either SPRINT pboot or native R parallelization*

The maximum number of resamples on HECTOR with the Golub dataset and median as the bootstrap statistic is between 75,000 and 80,000 pboot, for boot with multicore it is between 90,000 and 95,000. The serial version of boot however is able to handle more than 200,000 resamples, taking more than 4.5 hours to do so. This lower limit on the maximum number of resamples in the parallel implementations requires further investigation but is very likely due to the increased memory overhead on the master process. The multicore option fails with an out of memory error from the operating system while SPRINT pboot fails with an R message that the serialization is too large to store in a raw R vector. The master process in each case has objects for gathering the statistic results from all the worker processes. The large number of resamples combined with the number of processes means the resulting size of these objects on the master process and dealing with their distribution to the workers is the likely cause of these errors.

Table I gives the parallel speed up and efficiency for 75,000 resamples (i.e. close to the upper pboot limit) for boot with multicore and SPRINT pboot on multiple nodes. The serial version takes more than 100 minutes and when this is run with pboot on 512 nodes this is reduced to just over 1 minute. This close to 100 times speed up, indicates that, provided the aforementioned memory issues can be overcome then, pboot can perhaps offer a means to handle the expected volumes of next generation sequencing data in reasonable time scales.

With the current pboot implementation using R objects to handle the bootstrap resamples then two possible approaches for dealing with this limit are as follows. The first is to investigate R packages such as bigmemory [22] and the planned improvements in R Version 3 that enable the storage and manipulation of massive matrices. The second approach is to directly manipulate and manage the resamples on the master process more efficiently in C rather than R.

TABLE I. PARALLEL SPEED-UP AND EFFICIENCY FOR 75,000 RESAMPLES RELATIVE TO SERIAL R BOOT

| boot implementation, total number of processes, number of processes per node | Speed-Up | Efficiency |
|---|---|---|
| boot with multicore, on 32 processes, all on 1 node | 22.99 | 72% |
| SPRINT pboot, with 32 processes, 1 process per node | 26.44 | 83% |
| SPRINT pboot, with 256 processes, 1 process per node | 84.66 | 33% |
| SPRINT pboot, with 512 processes, 1 process per node | 97.20 | 19% |

VIII. CONCLUSIONS

A parallel bootstrap implementation for the SPRINT R package has been described. This has an almost identical interface and returns results in the identical R structure as the sequential boot version thus easing its inclusion in existing R scripts.

Depending on the complexity of the bootstrap statistic and the number of resamples, the SPRINT bootstrap, pboot, achieves speed ups of between 25 and just under 100 compared to the original serial code on a CRAY XE supercomputer. On this supercomputer, the multi-node performance of pboot compares favorably with a multicore parallel implementation of bootstrap in the R boot package itself. Further work is required to compare the performance of pboot with the cluster based parallel implementation that is also available in the R boot package.

Finally, it should be noted the performance of pboot could be further improved by wrapping different methods of random indices generation methods and calling them in parallel from individual workers.

Given the importance of next generation sequencing to life sciences and the volumes of data involved, more effort is required to ensure that parallel implementations of bootstrap in R can handle these. Dealing with larger data volumes and more resamples in a user-friendly manner and getting the best from the different computational architectures available, potentially with thousands of cores, requires more than the naïve approach to parallelization of farming out calculations. Such an approach clearly has limits and so more thought is required to get beyond these.

ACKNOWLEDGMENT

The SPRINT project has been funded by a Wellcome Trust grant [086696/Z/08/Z] and the Biotechnology and Biological Sciences Research Council (grant number BB/J019283/1). The work reported here was also funded under the HECToR distributed Computational Science and Engineering (CSE) Service operated by NAG Ltd. HECToR—A Research Councils UK High End Computing Service—is the UK's national supercomputing service, managed by EPSRC on behalf of the participating Research Councils. Its mission is to support capability science and engineering in UK academia. The HECToR supercomputers are managed by UoE HPCx Ltd and the CSE support Service is provided by NAG Ltd. http://www.hector.ac.uk